\documentclass[aps,prb,reprint,groupedaddress, floatfix]{revtex4-2}
\usepackage[hidelinks]{hyperref}
\usepackage{amsmath, amsfonts, amssymb}
\usepackage{bm}
\usepackage{natbib}
\usepackage{graphicx}
\usepackage{dcolumn}
\usepackage{color}
\usepackage{textcomp}
\usepackage{braket}

\allowdisplaybreaks
\newcommand{\overbar}[1]{\mkern 1.5mu\overline{\mkern-1.5mu#1\mkern-1.5mu}\mkern 1.5mu}
\DeclareMathOperator{\sech}{sech}

\begin{document}

\title{Ultrafast magnetization reversal in ferromagnetic spin-valves: an $s$-$d$ model perspective}

\author{Quentin Remy}
\email[Corresponding author: ]{quentinremy5@gmail.com}
\affiliation{Cavendish Laboratory, University of Cambridge, Cambridge, United Kingdom}
\affiliation{Université de Lorraine, CNRS, IJL, F-54000 Nancy, France}

\date{\today}

\begin{abstract}
We present an extension to simple $s$-$d$ models, aiming at simulating ultrafast magnetization dynamics and spin transport in metallic heterostructures. In particular, we consider an alternative spin dissipation channel due to a finite exchange splitting of the $s$ band. From this theory, we show three different mechanisms governing the dynamics of spin accumulation. On top of the already widely discussed ``$\text{-}\mathrm{d}M/\mathrm{d}t$" electron-magnon mechanism, we study the role of a dynamic change of exchange splitting (of conduction electrons) as well as the rotation of spins reflected at an interface with a ferromagnet. Finally, we use the presented theory to explain the recent observation of subpicosecond reversal of a ferromagnet in rare-earth free spin-valves. Our conclusion agrees with the one of reference \cite{Igarashi2023} favoring magnetization reversal due to the rotation of the spin polarization of a reflected spin current.
\end{abstract}

\maketitle

\section{\label{sec:intro}Introduction}

\footnotetext[1]{Certain works can be assigned to several categories, for instance extensions of the 3TM which are derived from a Boltzmann equation.}

Ever since the first investigations of ultrafast magnetization dynamics of metals \cite{Agranat1984, Vaterlaus1991, Beaurepaire1996, Scholl1997, Hohlfeld1997, Conrad1999}, numerous and diversified theoretical approaches have been attempted to understand its origin \cite{Scheid2022}. The proposed contributing microscopic mechanisms are typically classified as spin transport \cite{Battiato2010, Battiato2012a, Schellekens2013a, Graves2013, Wieczorek2015, Moisan2014, Dewhurst2018}, spin-flip scattering \cite{Koopmans2010a, Carva2011, Illg2013, Wieczorek2015, Baral2016} and magnon generation \cite{Carpene2008, Manchon2012, Haag2014, Tveten2015, Beens2022} processes. A sharp separation of the latter two mechanisms is however not always clear, mostly because the magnon generation mechanism is generally discussed in the framework of the electron-magnon interaction \cite{Haag2013, Haag2014} where the creation of a magnon always comes with a spin-flip, and also because both magnons and spin flips (Stoner excitations) are two specific cases of magnetic excitation \cite{Hertz1973a, Muller2019}. Assuming a clear separation between both types of excitation is possible, recent experimental results seem to confirm that both magnon generation and spin-flips are happening during ultrafast demagnetization (UDM) \cite{Turgut2016, Eich2017} and the final destination of angular momentum during magnetization quenching is the crystal lattice \cite{Dornes2019}. The generation of circularly polarized phonons was also observed for nickel \cite{Tauchert2022}. A transfer to the electromagnetic field has been estimated negligible \cite{Koopmans2003, DallaLonga2007} and a possible transient role of the orbital degree of freedom \cite{Zhang2021a} via, for instance, an increase of the orbital angular momentum \cite{Carpene2008, Chen2019}, still has not been observed to the best of our knowledge. Theoretical works however indicate that such an increase cannot be observed because orbital angular momentum in metals is transferred to the lattice with a characteristic time around one femtosecond \cite{Tows2015, Tows2023}. The computational frameworks that can incorporate some or all of these processes include the real-time time dependent density functional theory (rt-TDDFT) \cite{Krieger2015, Dewhurst2018, Chen2019, Acharya2020, Scheid, Elliott2022, Barros2022, Sharma2022}, a direct propagation of the system wave function with a parametrized many-body Hamiltonian \cite{Zhang2000, Tows2015, Tows2023, Stegmann2023}, the semiclassical Boltzmann equation \cite{Koopmans2005, Krauss2009, Battiato2010, Essert2011a, Battiato2012a, Mueller2013a, Illg2013, Mueller2013, Haag2014, Tveten2015, Nenno2018, Beens2020, Beens2022, Vollmar, Beens}, quantum kinetics \cite{Baral2015, Baral2016, Topler2020, Suresh2022}, molecular dynamics \cite{Ma2012, Pankratova2022}, the stochastic atomistic Landau-Lifshitz-Gilbert equation \cite{Kazantseva2008a, Zahn2021, Zahn2022, Pankratova2022}, the Landau-Lifshitz-Bloch equations \cite{Atxitia2010, Atxitia2017} as well as other more phenomenological parametrized models \cite{Rouzegar2022} such as the so-called three temperature model (3TM) \cite{Beaurepaire1996} and its various extensions \cite{Koopmans2010a, Manchon2012, Mueller2014, Kimling2017, Griepe, Note1}. We note that in parametrized approaches, parameters may be obtained from \textit{ab initio} calculations \cite{Lin2008, Carva2011, Maldonado2017, Scheid2019, Ritzmann2020}. In particular, recent works seem to validate the use of temperature based models for a description of magnetization dynamics \cite{Zahn2021, Zahn2022, Griepe, Pankratova2022} provided that one properly accounts for energy conservation and the temperature dependence of all parameters. Because there is so far no method that is able to completely solve the problem of UDM of metals for all relevant spatial and time scales, such simplified \textit{bath} models prove to be quite useful. 

Moreover, the field of ultrafast magnetization dynamics in metals is not limited to UDM. A plethora of consequent phenomena have been observed such as all-optical magnetization switching \cite{Stanciu2007, Radu2011, Mangin2014}, ultrafast spin injection \cite{Malinowski2008, Melnikov2011, Kampfrath2013, Choi2014, Choi2014a, Wieczorek2015, Choi2018a, Shin2018, Melnikov2022, Lichtenberg2022, Jimenez-Cavero2022, Rouzegar2022}, terahertz electromagnetic pulse emission \cite{Kampfrath2013, Seifert2016, Rouzegar2022, Seifert2022}, ultrafast spin transfer torque \cite{Schellekens2014a, Razdolski2017a, Lichtenberg2022} and magnetization reversal of ferromagnets using non-local transfer of spin \cite{Iihama2018, Remy2020, Igarashi2020, Iihama2021, Remy2023} or spin-orbit torque \cite{Jhuria2020}. We also note the apparent thermal/incoherent nature of these phenomena as experimentally highlighted by the fact that they do not intrinsically depend on the external energy stimulus used to trigger the dynamics \cite{Ostler2012, Bergeard2016, Chekhov2021, Remy2021}. Thus, the simplified parametrized models can be applied to study or predict more complex magnetization dynamics subsequent to UDM. It is also easier to enforce angular momentum conservation (although in a more phenomenological way), which is believed to not always be satisfied in more complicated approaches such as rt-TDDFT \cite{Simoni2022}.

In this work, we use a simple reservoir approach for the dynamics and transport of both energy and angular momentum, presented in Sec. \ref{sec:model}, which is suitable for the study of ultrafast thermal effects in metallic multilayers with possibly several magnetic layers. It is based on a $s$-$d$ model which can incorporate both transverse ($\sim$ magnons) and longitudinal (spin flips) magnetic excitations. In particular, this framework allows us to simulate a more complex spin accumulation dynamics which could explain recent experiments \cite{Igarashi2020, Igarashi2023}. Simulations of UDM and spin accumulation generation in systems with a single magnetic layer are briefly shown in Sec. \ref{sec:UDM}. Then, magnetization reversal in ferromagnetic spin-valves is discussed in Sec. \ref{sec:spin-valve}. We conclude and discuss possible improvements in Sec. \ref{sec:conclusion}.

\section{\label{sec:model}Model}

In order to provide some general context, we start the description of our model with a general model Hamiltonian $\mathcal{H}$ suitable for a $s$-$d$ model, in the absence of any external field and for a homogeneous material, where itinerant ($s$) electrons, phonons and localized ($d$) electrons are distinct quantities defined by their respective free (quasi)particle Hamiltonian terms $\mathcal{H}_e$, $\mathcal{H}_p$ and $\mathcal{H}_d$:
\begin{equation}\label{eq: H}
\mathcal{H} = \mathcal{H}_e+\mathcal{H}_p+\mathcal{H}_d+\mathcal{H}_{ep}+\mathcal{H}_{ed}
\end{equation}
together with two interaction terms $\mathcal{H}_{ep}$ and $\mathcal{H}_{ed}$ for the electron-phonon and $s$-$d$ interactions respectively. Two important approximations of this model are (i) the lack of hybridization between the $s$ and $d$ states and (ii) $d$ states do not contribute to (heat and charge) transport. The real impact of these approximations on the simulated dynamics is unclear since the final equations rely on parameters whose input values are often taken from experiments or \textit{ab initio} calculations that do not make these approximations. We write the $s$-$d$ interaction term as \cite{Cywinski2007,Gridnev2016,Remy2021}
\begin{equation}
\mathcal{H}_{ed} =-J\sum_i\left[ \dfrac{1}{2}\left( \hat{S}_i^+\hat{s}_i^-+\hat{S}_i^-\hat{s}_i^+ \right)+\hat{S}_i^z\hat{s}_i^z \right]
\end{equation}
where $J>0$ is the $s$-$d$ interaction constant for ferromagnetic coupling between $s$ and $d$ electrons, $\hat{\bm{S}}_i = (\hat{S}_i^x, \hat{S}_i^y, \hat{S}_i^z)$ are the spin operators for $d$ electrons localized on lattice site $i$, $\hat{\bm{s}}_i = \frac{1}{2}\hat{c}_{i}^\dagger\bm{\sigma}\hat{c}_{i}$ are the spin operators for $s$ electrons with $\hat{c}_{i\sigma}^\dagger$ and $\hat{c}_{i\sigma}$ the creation and annihilation operators for a Wannier state at lattice site $i$ and spin $\sigma$ and $\bm{\sigma} = (\sigma^x,\sigma^y,\sigma^z)$ is the vector of Pauli matrices. Superscripts $+$ and $-$ denote ladder operators as usual, for instance $\hat{S}_i^\pm = \hat{S}_i^x \pm \hat{S}_i^y$. The term with the $z$ components is treated in the mean field approximation \cite{Cywinski2007,Gridnev2016} and we write the average (over all quantum states and lattice sites) of the $z$ components of the itinerant and localized spins operators as $s^z$ and $S^z$ respectively. These two quantities contain all the angular momentum, per atom in units of $\hbar$, possessed by electrons within the mean field approximation. The Hamiltonian then becomes:
\begin{subequations}
\begin{align}
\mathcal{H} &= \mathcal{H}_e'+\mathcal{H}_p+\mathcal{H}_d'+\mathcal{H}_{ep}+\mathcal{H}_{ed}^\pm\\
\mathcal{H}_e' &= \mathcal{H}_e-JS^z\sum_{\bm{k}}\dfrac{1}{2}\left( \hat{c}_{\bm{k}\uparrow}^\dagger\hat{c}_{\bm{k}\uparrow} - \hat{c}_{\bm{k}\downarrow}^\dagger\hat{c}_{\bm{k}\downarrow} \right) \label{eq: He'}\\
\mathcal{H}_d' &=\mathcal{H}_d-Js^z\sum_i\hat{S}_i^z\label{eq: Hd'}\\
\mathcal{H}_{ed}^\pm &= -J\sum_i\dfrac{1}{2}\left( \hat{S}_i^+\hat{s}_i^-+\hat{S}_i^-\hat{s}_i^+ \right)
\end{align}
\end{subequations}
Where we wrote the mean field term felt by the $s$ electrons in the Bloch representation (with states indexed by a wavevector $\bm{k}$).

From this Hamiltonian, the system dynamics is obtained in the framework of perturbation theory: the interaction terms $\mathcal{H}_{ep}$ and $\mathcal{H}_{ed}^\pm$ are assumed to be much smaller than the free particle terms. The interaction terms are then used to determine exchange of energy (as well as momentum and angular momentum) while the free particle terms are used for conservation of energy (as well as momentum and angular momentum). Notably, the interaction energies are neglected in the conservation of energy, which is not true in general \cite{Remy2021}, but is valid whenever perturbation theory can be applied.

We first discuss how the free particle terms are treated. The itinerant free electron term $\mathcal{H}_e'$ describes renormalized (for instance due to the electron-phonon and electron-electron interaction) Bloch states which are assumed to be characterized by a thermal distribution for each (pure) spin state. The electronic temperature $T_e$ is assumed to be the same for both spin species $\sigma = \pm 1/2$ while the chemical potential $\mu_\sigma$ and the density of states $D_\sigma$ are different for different spins. The mean field term in Eq.(\ref{eq: He'}) adds an exchange splitting to the $s$ band. Noting $E_\sigma^0$ the lowest energy of the $s$ band, the exchange splitting is seen to be $E_\uparrow^0-E_\downarrow^0=-JS^z$. The free phonon term $\mathcal{H}_p$ represents renormalized phonons \cite{Mahan2000} described as Debye phonons in equilibrium at a temperature $T_p$. The localized electrons term $\mathcal{H}_d$ can contain contributions due to $d$-$d$ exchange interaction, magnetocrytalline anisotropy and dipolar interaction \cite{Tveten2015}. It is also treated within the mean field approximation \cite{Beens2020}. We consider for simplicity the case where the spin quantum number $S$ of the $d$ electrons is 1/2. The total mean field, including the one generated by the $s$-$d$ interaction (second term of Eq.(\ref{eq: Hd'})), induces an energy splitting $\Delta = 2mk_BT_C$ between the $d$ electrons energy levels where $m=-2S^z$ is the magnetization of the $d$ electrons normalized to its zero temperature value and $T_C$ is the Curie temperature of the ferromagnetic layer. We note that the $d$ electrons are not assumed to be in internal equilibrium (in the $d$ electrons bath itself) for a general value of $S$, but for $S=1/2$, in this mean field approximation, there is no difference between internal equilibrium and out-of-equilibrium \cite{Remy2023}.

The electron-phonon term $\mathcal{H}_{ep}$ is treated in perturbation theory via Fermi's Golden rule \cite{Allen1987,Mahan2000,Chen2015}. We consider the usual high temperature case where the energy transfer between electrons and phonons is found to be $g_{ep}(T_e-T_p)$ with $g_{ep}$ the electron-phonon coupling considered as temperature independent. We also consider that this term induces angular momentum transfer between the $s$ electrons and the lattice and which is phenomenologically given by, following references \cite{Cywinski2007, Gridnev2016}, $(s^z-s^z_{\text{ie}})/\tau_s$ with $\tau_s$ a spin relaxation time. The instantaneous equilibrium $s$ electrons spin polarization $s^z_{\text{ie}}$ is defined such that $s^z-s^z_{\text{ie}}$ is the excess of spin due to the out-of-equilibrium state of the itinerant electrons \cite{Tveten2015}. It is discussed bellow. The exchange of energy due to the $s$-$d$ interaction term $\mathcal{H}_{ed}^\pm$ is obtained via a slightly generalized version of Fermi's Golden rule \cite{Blum2012, Cywinski2007, Beens2020} which leads to a typical two-level dynamics \cite{Koopmans2010a, Nieves2014, Beens2020}:
\begin{equation}\label{eq: dmdt}
\dfrac{dm}{dt} = \dfrac{1}{\tau_m}\left( m-\dfrac{\Delta\mu}{2k_BT_C} \right)\left[ 1-m\coth\left( \dfrac{2mk_BT_C-\Delta\mu}{2k_BT_e} \right) \right]
\end{equation}
Where $\tau_m$ is the characteristic time for angular momentum transfer from $d$ to $s$ electrons, which needs to be taken as an additional parameter \cite{Beens2020, Remy2021}, and $\Delta\mu = \mu_\uparrow-\mu_\downarrow$ is the spin accumulation of the $s$ electrons.

The equations governing the dynamics of the system come from the fact that conservation equations must be fulfilled while transfers happen as given by interactions, which we just discussed. We only focus on energy and angular momentum conservation equations, and discard effects appearing when one also considers charge \cite{Kimling2017} and momentum \cite{Chen2006} conservation, as we wish to discuss new effects arising from angular momentum conservation driven by energy transfer. The energy conservation equation for the total Hamiltonian Eq.(\ref{eq: H}) reads:
\begin{equation}\label{eq: conv heat}
\dfrac{\partial}{\partial t}\left( \dfrac{\gamma}{2}T_e^2 + C_p T_p - \rho m\dfrac{\Delta}{2} \right) + \bm{\nabla}\cdot \left( \bm{Q}_e + \bm{Q}_p \right) = 0
\end{equation}
Where we chose the middle of both $d$ electrons levels as the reference energy for $d$ electrons, $\rho$ is the number of atoms per unit volume, $\gamma T_e$ is the standard expression for a free electron gas volumetric heat capacity, $C_p$ is the phonon volumetric heat capacity and $\bm{Q}_e$ and $\bm{Q}_p$ are the electronic and phononic heat current densities respectively. The phononic heat current is given by the standard Fourier's law $\bm{Q}_p = -\kappa_p \bm{\nabla}T_p$, with $\kappa_p$ the phonon heat conductivity, while the electronic heat current is given by $\bm{Q}_e = -\kappa_e (T_e/T_p) \bm{\nabla}T_e$ with $\kappa_e$ the equilibrium (when the electronic plus phononic system is in equilibrium) electronic heat conductivity. Such a description of the energy flow within electrons and phonons was recently used to successfully describe the ultrafast strain dynamics of heterostructures similar to the ones considered in our work \cite{Pudell2020}. The resulting heat equations for the system, modeled as one dimensional in the thin film limit, are:
\begin{subequations}\label{eq: T}
\begin{align}
\begin{split}\label{eq: Te}
\gamma T_e \dfrac{\partial T_e}{\partial t} &= \dfrac{\partial}{\partial z}\left( \kappa_e \dfrac{T_e}{T_p} \dfrac{\partial T_e}{\partial z} \right) - g_{ep}\left( T_e - T_p \right) \\
&+2 \rho mk_BT_C\dfrac{dm}{dt} + S(z,t)
\end{split}
\\[2ex]
\begin{split}
C_p \dfrac{\partial T_p}{\partial t} &= \kappa_p \dfrac{\partial^2 T_p}{\partial z^2} + g_{ep}\left( T_e - T_p \right)
\end{split}
\end{align}
\end{subequations}
The energy dynamics for the $d$ electrons is given by Eq.(\ref{eq: dmdt}) since there is a one-to-one relationship between the energy density $-\rho m\Delta/2$ and the absolute value of magnetization $|m|$ in this model. The magnetization dependent term in Eq.(\ref{eq: Te}) \cite{Gridnev2018, Griepe} comes from the requirement that Eq.(\ref{eq: conv heat}) must be fulfilled. The last term in Eq.(\ref{eq: Te}) is due to energy transfer from an external laser pulse as computed in reference \cite{Chavazas2022} where it is argued that energy conservation in the total system, including the electromagnetic field, is significantly broken. Satisfying conservation of energy when the electromagnetic field is included is however irrelevant to the phenomena discussed in this work and beyond the scope of this work \cite{Philip}.

The angular momentum conservation equation is:
\begin{equation}\label{eq: sz}
\dfrac{\partial}{\partial t}\left( s^z+S^z+S_p^z \right) + \bm{\nabla}\cdot \bm{J}_s = 0 .
\end{equation}
Similarly to energy conservation, quantities appearing after the time derivative operator are intensive quantities. In this case we take them as angular momentum (or spin polarization for the electrons) per atom in units of $\hbar$ to have notations consistent with the previously introduced averaged spins. $S_p^z$ refers to the angular momentum dissipated in the lattice which, according to our previous discussion, satisfies $\partial S_p^z/\partial t = (s^z-s^z_{\text{ie}})/\tau_s$. $\bm{J}_s$ is the spin current density and we neglect a contribution to this current density due to angular momentum transport in the localized $d$ electrons \cite{Cornelissen2016b, Beens2022} or in phonons \cite{Ruckriegel2020}. Moreover, we assume that spin transport only happens via conduction electrons close to the Fermi level \cite{Beens} and so depends on the spin accumulation only \cite{Beens2020, Lichtenberg2022, Rouzegar2022}. We will detail the spin current term when we use it in section \ref{sec:spin-valve}.

To close our system of equations, and because spin transport depends on the spin accumulation, we need to rewrite Eq.(\ref{eq: sz}) in terms of $\Delta\mu$. The relation we need is \cite{Mueller2013, Tveten2015}
\begin{equation}\label{eq: s_deltamu}
s^z-s^z_{\text{ie}} = \overbar{D}\left( \Delta \mu -\delta \right),
\end{equation}
Which is valid to first order in both the spin accumulation $\Delta \mu$ and the change of exchange splitting $\delta = -J(S^z-S^z_{\text{ie}})$ with $S^z_{\text{ie}}$ the instantaneous equilibrium value of $S^z$. $N\overbar{D}=D_\uparrow(\varepsilon_F)D_\downarrow(\varepsilon_F)/(D_\uparrow(\varepsilon_F)+D_\downarrow(\varepsilon_F))$, with $\varepsilon_F$ the equilibrium Fermi level and $N$ the total number of $s$ electrons, and it is taken as a parameter \cite{Tveten2015,Beens2020}. A derivation of Eq.(\ref{eq: s_deltamu}) is provided in Appendix \ref{sec: A3}. This equation also allows us to consider the change of exchange splitting in the conduction electrons which was argued to be fundamental to describe ultrafast magnetization dynamics of itinerant ferromagnets \cite{Schellekens2013, Mueller2013} and later considered in both itinerant and localized electrons \cite{Tveten2015}. We also need to calculate the instantaneous equilibrium values of $s^z$ and $S^z$. For $s^z_{\text{ie}}$, we follow the argumentation of Gridnev \cite{Gridnev2016}, simplified to the case of ferromagnets:
\begin{equation}\label{eq: chi}
s^z_{\text{ie}}(t) = \chi S^z(t)
\end{equation}
With $\chi$ a spin susceptibility. When ferromagnetic order only arises due to the $s$-$d$ interaction, $\chi = 4k_BT_C/J$. The choice of $S^z_{\text{ie}}(t)$ is more complicated. The naive case where $S^z_{\text{ie}}(t)$ is given by its equilibrium value for a temperature given by the electronic temperature at the instant of interest $T_e(t)$ would be unphysical. In particular, it would lead to a fast change (even a discontinuity in the mean field approximation) of the slope of the spin accumulation dynamics curve when the electronic temperature crosses the Curie temperature. Rather, we follow the physics of out-of-equilibrium spin relaxation, where the dynamics of localized spins for $S=1/2$ is governed by an equation with the following form \cite{Koopmans2010a,Nieves2014, Beens2020, Remy2023}:
\begin{subequations}
\begin{align}\label{eq:better_self_c_bloch}
\dfrac{dm}{dt}_{|\text{relaxation}} &= -\dfrac{m(t)-m_{\text{ie}}(t)}{\tau(t)}\\ \label{eq: mie}
m_{\text{ie}}(t) &= -2S^z_{\text{ie}}(t) \equiv \tanh\left( \dfrac{2k_BT_Cm(t)-\Delta \mu(t)}{2k_BT_e(t)} \right)
\end{align}
\end{subequations}
Where $\tau(t)$ is a characteristic time that depends on time and, in general, $\Delta \mu$ represents an energy splitting due to an external (to the $d$ electrons subsystem) source of angular momentum. For instance, for the $s$-$d$ model \cite{Beens2020} fundamentally describing, at each instant $t$, spin relaxation of $d$ electrons in the thermal bath of $s$ electrons
\begin{equation}
\tau(t) = \dfrac{m_{\text{ie}}(t)\tau_m}{m(t)-\Delta \mu(t)/(2k_BT_C)},
\end{equation}
While for Elliott-Yafet scattering as computed by Koopmans \textit{et al.} \cite{Koopmans2010a}
\begin{equation}
\tau(t) = \dfrac{m_{\text{ie}}(t)T_C}{m(t)RT_p(t)}
\end{equation}
With $R$ the demagnetization rate in the Elliot-Yafet model \cite{Koopmans2010a} and an external source of angular momentum can also be considered \cite{Cornelissen2016}. Eq.(\ref{eq:better_self_c_bloch}) is more general than a Bloch equation, and even more general than (the longitudinal term of) the self-consistent Bloch equation \cite{Xu2012, Xu2013} because $\tau$ depends on time in a complicated way. Here we keep $\tau(t) = \tau_s$ which is consistent with the naive description of spin dissipation Eq.(\ref{eq: sz}) \cite{Cywinski2007,Gridnev2016} and the self-consistent Bloch equation \cite{Xu2012, Xu2013}.

Using equations (\ref{eq: s_deltamu}) and (\ref{eq: chi}), Eq.(\ref{eq: sz}) becomes \cite{Remy2021}:
\begin{widetext}
\begin{equation}\label{eq:dmu}
\dfrac{d\Delta \mu}{dt} = \left( \dfrac{S}{\overbar{D}}(1+\chi) +JS \right)\dfrac{dm}{dt} - JS\dfrac{dm_{\text{ie}}}{dt} - \dfrac{\bm{\nabla}\cdot \bm{J}_s}{\overbar{D}} - \dfrac{\Delta \mu}{\tau_s} + JS \dfrac{m-m_{\text{ie}}}{\tau_s}
\end{equation}
\end{widetext}
Which together with equations (\ref{eq: dmdt}), (\ref{eq: T}) and (\ref{eq: mie}) form the set of equations we wish to solve to obtain the dynamics of $T_e$, $T_p$, $m$ and $\Delta \mu$. Other quantities such as the exchange splitting or the total spin polarization in the electronic subsystem can be obtained from the latter four quantities. Eq.(\ref{eq:dmu}) is valid for any value of $S$ but more equations are then needed to calculate the dynamics of $m$ and $m_{\text{ie}}$. Eq.(\ref{eq:dmu}) generalizes previous approaches \cite{Gridnev2016, Beens2020} mainly because it includes a dynamic exchange splitting of the $s$ electrons (terms proportional to $JS$). It also includes an equilibrium spin polarization of these electrons which was not in the model of Beens \textit{et al.} \cite{Beens2020}. This dynamic exchange splitting was considered before by Tveten \textit{et al.} \cite{Tveten2015} and we note a similarity between the form of our equations (\ref{eq: dmdt}) and (\ref{eq:dmu}) and equations (5) and (4) of reference \cite{Tveten2015} if one replaces $m$ and $m_{\text{ie}}$ by the out-of-equilibrium and equilibrium (Bose-Einstein) magnon distribution respectively. The system of equations of Tveten \textit{et al.} is however much more complicated than ours to solve and it is also not clear whether the magnonic description of the magnetization dynamics via the Holstein-Primakoff expansion is valid especially since we wish to model situations where magnetization can be fully quenched or even reversed \cite{Bajpai2021}. Finally, a new mechanism appearing in our approach is a spin dissipation in the lattice due to a non-zero value of the out-of-equilibrium magnetization $m-m_{\text{ie}}$ of $d$ electrons. All previous works so far (in this framework) have, as far as we know, only been considering a spin dissipation due to the presence of a spin accumulation as defined above. Within the context of the derivation of Eq.(\ref{eq: s_deltamu}), this means that we consider spin relaxation in the lattice due to a spin non-equilibrium in $s$ electrons close to the Fermi level (the spin accumulation dissipation term $- \Delta \mu/\tau_s$) as well as all the other ones (the dynamic exchange splitting dissipation term $JS (m-m_{\text{ie}})/\tau_s$). This additional contribution comes from the fact that all electronic states are considered to contribute to the spin dissipation when the phenomenological term $\partial S_p^z/\partial t = (s^z-s^z_{\text{ie}})/\tau_s$ \cite{Cywinski2007,Gridnev2016} is assumed. It was argued, in a different framework, that a relaxation-time approximation can be used to simulate UDM, with a relaxation-time identical for all electronic states \cite{Vollmar}. One potential interesting consequence of this additional term is that it can change the sign of the spin accumulation as compared to what can be expected from the usual ``$\text{-}\mathrm{d}M/\mathrm{d}t$" law for the spin generation rate (first term on the right hand side of Eq.(\ref{eq:dmu})) since when $dm/dt$ is negative, $m-m_{\text{ie}}$ is usually positive (this depends on the dynamics of the electronic temperature and the spin accumulation). Also note that, even though equation (\ref{eq: dmdt}) can be written as equation (\ref{eq:better_self_c_bloch}), $dm/dt$ is not proportional (as a function of time) to $m-m_{\text{ie}}$ due to the complex time dependence of $\tau$.

In order to facilitate the numerical implementation of this model and to explain its limitations, it is instructive to expand Eq.(\ref{eq:dmu}) using Eq.(\ref{eq: mie}):
\begin{widetext}
\begin{subequations}
\begin{gather}
\begin{split}\label{eq:final_dmu}
\dfrac{d\Delta \mu}{dt} = \dfrac{1}{1-\dfrac{JS\eta(t)}{2k_BT_e(t)}} \times \left[ \left( \dfrac{S}{\overbar{D}}\left(1+\chi\right)+JS\left( 1-\eta(t)\dfrac{T_C}{T_e(t)} \right) \right) \dfrac{dm}{dt} \right.
\\[2ex]
+ \left.\dfrac{JS\eta(t)\xi(t)}{T_e(t)} \dfrac{dT_e}{dt} - \dfrac{\bm{\nabla}\cdot \bm{J}_s}{\overbar{D}} - \dfrac{\Delta \mu(t)}{\tau_s} + \alpha JS \dfrac{m(t)-m_{\text{ie}}(t)}{\tau_s}\right]
\end{split}
\\[2ex]
\begin{split}
\eta(t) \equiv \sech^2\left( \xi(t) \right) \quad\text{;}\quad \xi(t) \equiv \dfrac{2k_BT_Cm(t)-\Delta \mu (t)}{2k_BT_e(t)}
\end{split}
\end{gather}
\end{subequations}
\end{widetext}
Additional terms appears due to the dependence of $m_{\text{ie}}$ on $T_e$, $m$ and $\Delta \mu$. These terms are proportional to the quantity $\eta$ which becomes sizable when $\xi$ is small i.e. at large electronic temperature and for values of magnetization where effects such as spin cooling and spin heating \cite{Remy2023} are expected to become significant (when $2k_BT_Cm(t) \sim \Delta \mu (t)$). Moreover, we notice that the change of spin accumulation diverges when $(JS\eta(t))/2k_BT_e(t)) \sim 1$, which may happen even when $\xi$ is large depending on the complexity of the dynamics of $T_e$, $m$ and $\Delta \mu$. This divergence happens when the first order expansions of the density of states of $s$ electrons leading to equation (\ref{eq: s_deltamu}) are no longer valid approximations. These approximations are valid as long as $J \ll 1/\overbar{D}$ (see Appendix \ref{sec: A3}). This is consistent with the fact \cite{Stiles2002, Beens2020} that the width of the conduction band is usually larger than $JS$. The terms proportional to $J$ should then be treated as a correction to the model of Beens \textit{et al.} \cite{Beens2020} however we will show that they can significantly change the magnetization dynamics when $dm/dt \sim 0$ in the presence of an external source of spin accumulation. Equation (\ref{eq:final_dmu}) not only shows that the spin accumulation and the spin current are not always proportional to $-dm/dt$ \cite{Beens2022} but the spin generation rate itself \cite{Choi2014} is also not always proportional to $-dm/dt$ due to the dynamic exchange splitting. We introduced a parameter $\alpha$ which is equal to 1 in our model but we will also set it to 0 in the next section to see the effect of the corresponding term in equation (\ref{eq:final_dmu}).


Because both $s$ and $d$ electrons carry angular momentum in this model, even at equilibrium, the question arises as to which quantity is measured in experiments. Indeed, it is not clear whether an optical probe measuring magnetization via magneto-optical effects would be as sensitive to both kinds of electrons or not. Here we will assume that the experimentally measured quantity is proportional to the total spin angular momentum of the system as we are only interested in qualitative modeling. Then a magnetic signal will be proportional to
\begin{equation}
\begin{split}
-S^z_{\text{tot}} &\equiv -\left(S^z+s^z\right) \\
&= S\left( m\left( 1+\chi \right) +J\overbar{D}\left( m-m_{\text{ie}} \right)\right)-\overbar{D}\Delta\mu
\end{split}
\end{equation}
Showing that even under our simple approximation Eq.(\ref{eq: chi}), the signal will not be proportional to the $d$ electrons normalized magnetization $m$ due to a non zero spin accumulation \cite{Beens2020} and dynamic exchange splitting. All the data we plot is normalized by the equilibrium spin angular momentum $-Sm(1+\chi)$ and we note the corresponding normalized magnetization $m_{\text{tot}}$.

The systems we wish to simulate are actually multilayers and are thus not homogeneous. We thus assume as usual that all the previous equations are valid within each layer of the multilayer. We solve the conservation of energy equation (\ref{eq: conv heat}) by discretizing each layer and we use appropriate boundary conditions for each interface \cite{Kimling2017, Remy2021}. For the conservation of angular momentum, we assume that magnetization is constant along the thickness of each magnetic layer and use the average of the electronic temperature in the corresponding layer as the input temperature appearing in equations (\ref{eq: dmdt}) and (\ref{eq:dmu}). Thus the term $2mk_BT_Cdm/dt$ in Eq.(\ref{eq: conv heat}) is identical for all depths in a given magnetic layer. Going beyond this approximation would require a generalization of this $s$-$d$ model to include either a direct $d$-$d$ coupling \cite{Beens2019a} or indirect $s$-$d$ coupling between neighboring atomic layer. Such approach is beyond the scope of this work.

\section{\label{sec:UDM}Ultrafast demagnetization of a single layer}

We first present results for a multilayer structure with a single ferromagnetic layer, namely Sapphire(Substrate) / Ta(5) / Pt(4) / [Co/Pt](3.2) / Ta(5) similarly to reference \cite{Igarashi2023} where numbers between brackets are thicknesses in nm and [Co/Pt] is a ferromagnetic multilayer which is simulated as an effectively homogeneous medium. We neglect spin transport in this case. We use a 50 fs (gaussian) laser pulse, with normal incidence (coming from the sample side i.e. the air/Ta interface) and 800 nm central wavelength to bring the system out of equilibrium. We study the dynamics of various quantities as a function of time delay with respect to the time instant where the position of the center of the laser pulse is at the air/Ta interface. The sample is initially at equilibrium with room temperature chosen to be 300 K. The parameters entering the energy conservation equation are taken from our previous works \cite{Igarashi2020, Remy2021, Remy2023} and we choose $\rho = 7.5 \times 10^{28} ~ \text{m}^{-3}$ which lies between the values of pure Co ($\sim 9 \times 10^{28} ~ \text{m}^{-3}$) and Pt ($\sim 6.5 \times 10^{28} ~ \text{m}^{-3}$). For the angular momentum parameters we choose $\tau_m =$ 100 fs, $\tau_s =$ 20 fs and $1/\overbar{D} =$ 1 eV as in reference \cite{Beens2020}. We choose to study $\chi =$ 1 such that equation (\ref{eq:dmu}) reduces to equation (6) of reference \cite{Beens2020} when $J =$ 0 eV, since we have $S = 1/2$. These parameters will be kept in the rest of this work. In this section, we choose $J =$ 0 or 0.1 eV and $T_C =$ 700K. We will also study the effect of the newly proposed spin dissipation channel by using $\alpha =$ 0 or 1. All the reported fluences are external fluences i.e. not the absorbed ones.

\begin{figure*}[t!]
\centering
\includegraphics[scale=1]{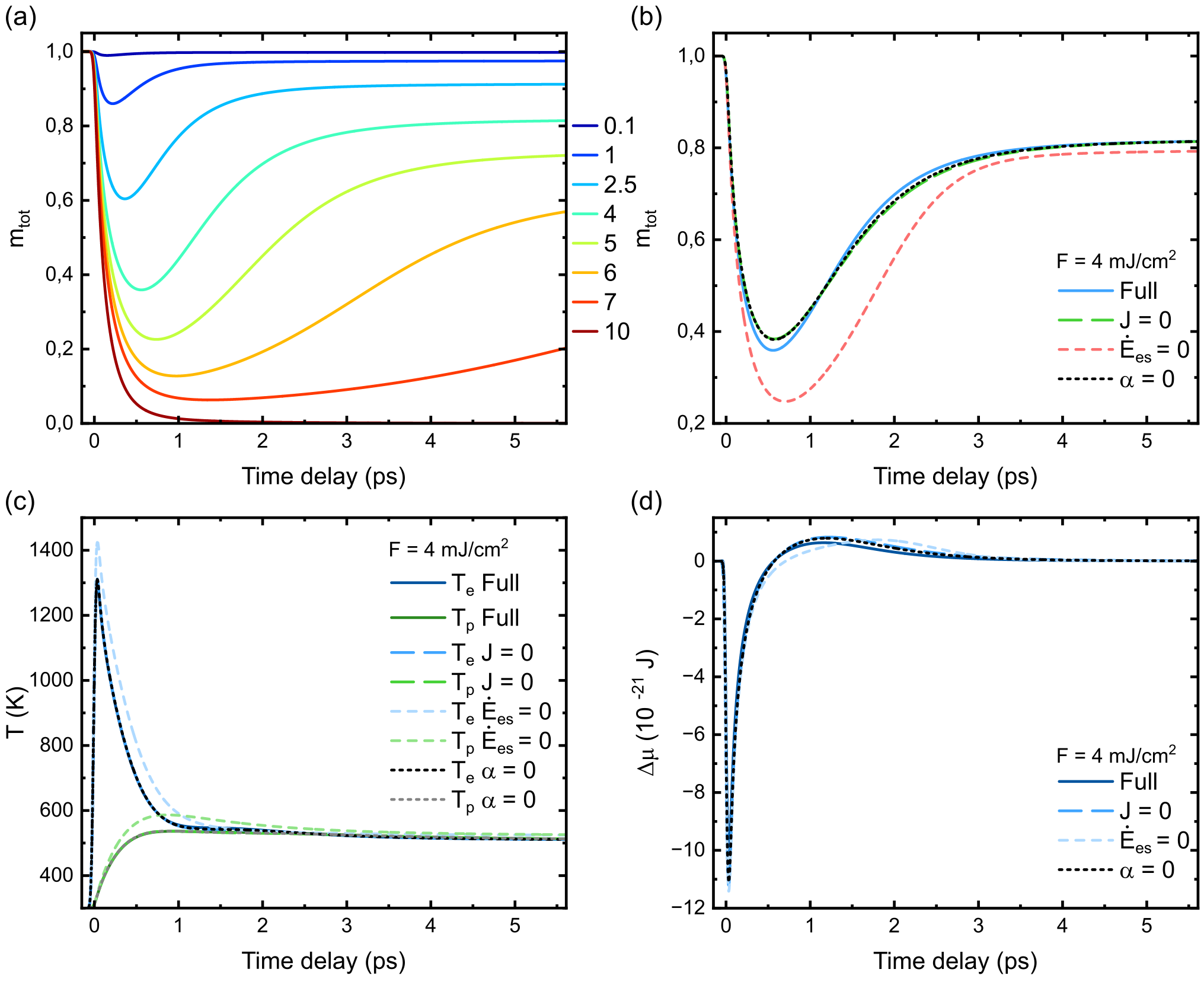}
\caption[Single_layer]{(a) Normalized magnetization $m_{\text{tot}}$ of the [Co/Pt] multilayer as a function of time delay for various fluences as noted on the right of the plot (in $\text{mJ/cm}^\text{2}$). (b) Comparison of the magnetization dynamics for a fluence of 4 $\text{mJ/cm}^\text{2}$ and different models; ``Full" corresponds to the full equation (\ref{eq:final_dmu}) with $J =$ 0.1 eV and $\alpha =$ 1; ``$J =$ 0" is obtained by setting $J =$ 0 eV; ``$\dot{E}_{\text{es}} =$ 0" is obtained by neglecting the magnetization dependent term in equation (\ref{eq: Te}); ``$\alpha =$ 0" is obtained by setting $\alpha$ to zero. (c) Same comparison as in (b) but for the electron and phonon temperature as indicated. (d) Same comparison as (b) but for the spin accumulation as indicated.}\label{fig:single_layer}
\end{figure*}

Fig. \ref{fig:single_layer}(a) shows the magnetization dynamics of the [Co/Pt] multilayer for various fluences. We recover the standard behavior, with an UDM followed by a ``fast" recovery (where spins and electrons do not form an equilibrated sub-bath of the system) and a ``slow" recovery (where spins and electrons are equilibrated and the dynamics is driven by the phonon temperature dynamics via heat dissipation in the substrate). At higher fluences, we also recover the so-called Critical Slowing Down (CSD) of magnetization dynamics \cite{Kazantseva2008a, Koopmans2010a, Atxitia2010, Manchon2012, Xu2012, Remy2023}. In Fig. \ref{fig:single_layer}(b), (c) and (d), we plot the normalized magnetization, temperatures and spin accumulation respectively, for four different models. All four models lead to qualitatively identical dynamics. A quantitative difference is observed for fluences such that there is a significant quenching of magnetization and yet at the same time a significant remagnetization. All four models lead to almost identical dynamics when there is a significant CSD (not shown). Neglecting the dynamic exchange splitting can change the value of the normalized magnetization by a few percent during the UDM and ``fast" recovery phases. By only turning off the spin dissipation channel due to the dynamic exchange splitting, we recover almost the same dynamics as when we completely turn off the dynamic exchange splitting, indicating that the dissipation part dominates the dynamics induced by the dynamic exchange splitting. A bigger effect on the magnetization dynamics is obtained by turning off transfer of energy from $d$ to $s$ electrons. This is because it modifies the temperature dynamics as shown in Fig. \ref{fig:single_layer}(c).

Overall, we do not observe any drastic difference between the model of Beens \textit{et al.} \cite{Beens2020} and ours, even in the high fluence limit. The reason is that, when $dm/dt \sim 0$ and so when terms proportional to $J$ could dominate, the temperature dynamics is already much slower and the self-consistency of our system of equations forces $d\Delta \mu/dt \sim 0$.  The situation will be completely different in the next section where an external source of angular momentum (due to an additional ferromagnetic layer) can also drive the spin accumulation dynamics.

\section{\label{sec:spin-valve}Subpicosecond magnetization switching in ferromagnetic spin-valves}

The aim of this section is to study potential mechanisms that can lead to the subpicosecond magnetization switching of ferromagnets observed in reference \cite{Igarashi2023}. In this case, the system is Sapphire(Substrate) / Ta(5) / Pt(4) / [Co/Pt](7) / Cu(10) / [Co/Pt](3.2) / Ta(5) where the first [Co/Pt] multilayer (7 nm thick) is referred to as the ``Reference" layer and the second [Co/Pt] multilayer (3.2 nm thick) is referred to as the ``Free" layer. A magnetic configuration of the system where the magnetizations of each ferromagnetic multilayer are parallel is noted ``P" and if they are antiparallel, we note it ``AP". The main result of reference \cite{Igarashi2023} was to show that upon a single femtosecond laser pulse irradiation of the sample, the free layer can reverse its magnetization. For such a thickness of Cu, this can happen whether the sample is initially prepared in either a P or an AP configuration. However, less fluence is systematically required to reverse the magnetization of the free layer from a P configuration compared to the AP configuration. Also, the dynamics of the reversal was measured for an initial P configuration and the free layer magnetization was seen to cross zero before the reference layer starts to remagnetize. It was shown in reference \cite{Igarashi2023} that the model of Beens \textit{et al.} \cite{Beens2020} can reproduce the magnetization reversal from the P configuration, but magnetization crosses zero during the remagnetization of the reference layer and it can therefore not explain the experimental measurements of reference \cite{Igarashi2023}. Igarashi \textit{et al.} \cite{Igarashi2023} therefore concluded that another mechanism has to come into play and suggest a mechanism where a spin current generated by the demagnetization of the free layer is reflected at the Cu/Reference layer interface, and upon this reflection, the polarization of the spin current can be rotated. This phenomenon is already well-known in the context of spin transfer torque in non-colinear spin configurations \cite{Stiles2002}. We propose here to make a simplified model of this mechanism to show that it can explain the qualitative behavior observed in reference \cite{Igarashi2023}.

\begin{figure*}[t!]
\centering
\includegraphics[scale=1]{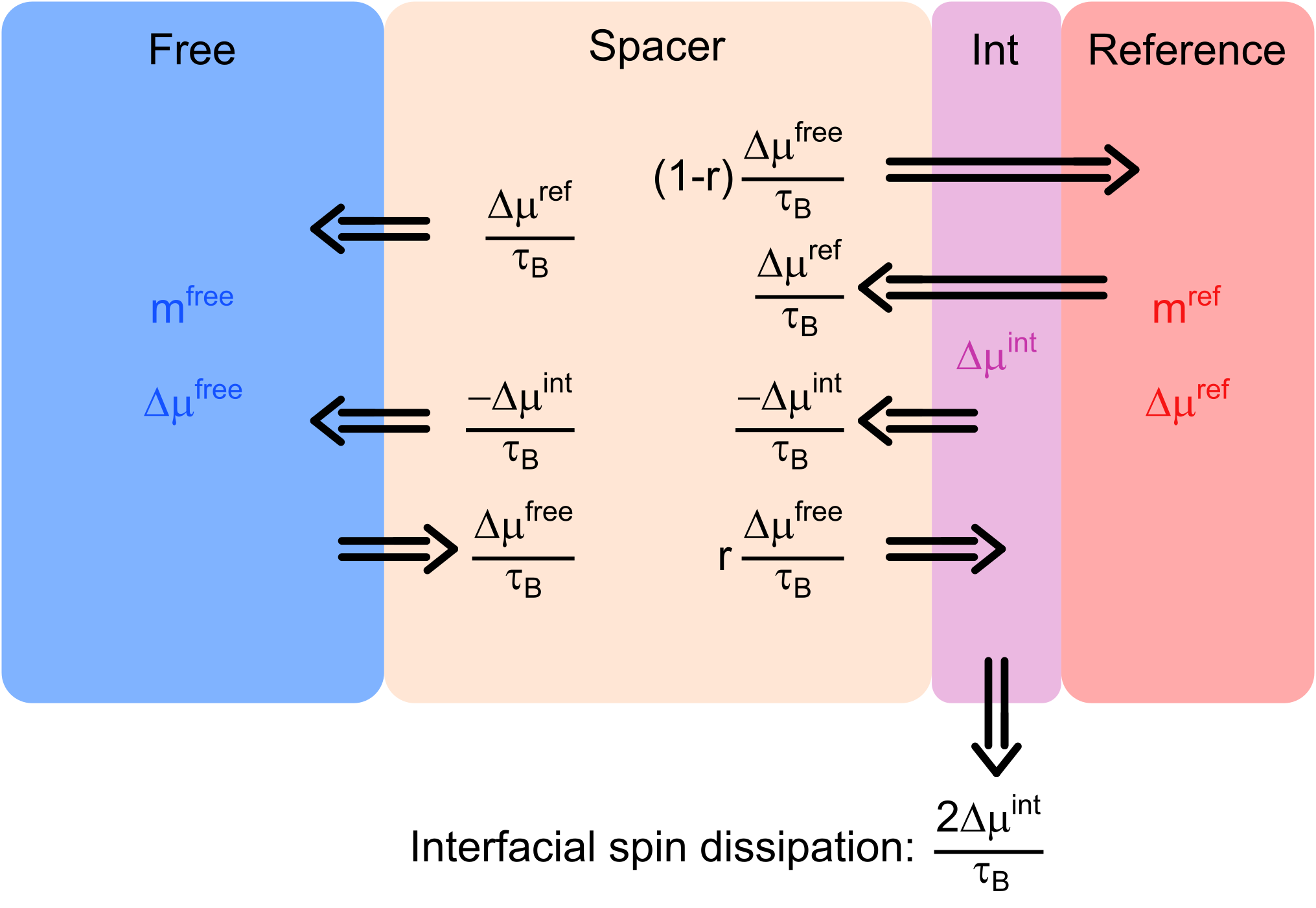}
\caption[mu_s_int]{Schematic description of the ballistic spin transport model. Black arrows represent the various spin currents considered at each interface. The corresponding terms in the spin transport equations (\ref{eq:js}) and (\ref{eq:mu_s_int_dyn}) are shown next to each arrow. ``Spacer'' refers to the Cu spacer layer and ``Int" to the Cu/Reference layer interface.}\label{fig:mu_s_int}
\end{figure*}

The parameters that we use are the same as in the previous section. The only exception is that we take a Curie temperature of 500 Kelvin and $J =$ 0.05 eV for the free layer. We now need to include spin transport and so we no longer neglect the spin current term of Eq.(\ref{eq:dmu}) unless it is explicitly stated. We make use of the now well-established result that spin currents are proportional to the spin accumulation when a ferromagnetic layer is in contact with a good spin sink \cite{Lichtenberg2022, Beens2022, Beens, Rouzegar2022}. In our case, considering a given ferromagnetic layer, we assume ballistic spin transport in the Cu spacer layer and this spin sink is the other ferromagnetic layer. For a single ferromagnet/Cu interface, we then assume that the spin current exiting the ferromagnet obeys the following equation \cite{Beens2020, Lichtenberg2022}:
\begin{equation}
\dfrac{\bm{\nabla}\cdot \bm{J}_s}{\overbar{D}} = \dfrac{\Delta \mu}{\tau_B}
\end{equation}
Where, following Beens \textit{et al.} \cite{Beens2020} we take $\tau_B =$ 10 fs for our 10 nm of Cu. Our approach of the simulation of ballistic spin transport is similar to Ref. \cite{Beens2020}. We have a single (i.e. depth independent) $d$ electrons magnetization $m^{\text{free}}$ and $m^{\text{ref}}$ in the free and reference layers respectively, as well as depth independent spin accumulations $\Delta\mu^{\text{free}}$ and $\Delta\mu^{\text{ref}}$. The main difference is that we consider that the Cu/Reference layer interface has a separate spin accumulation $\Delta\mu^{\text{int}}$. The situation is summarized in Fig.\ref{fig:mu_s_int}.
Because we want to simulate reflection at the Cu/Reference layer interface, we introduce a parameter $r \in [0,1]$ that quantifies the amount of spin that is reflected from the interface. $r$ times the spin current coming from the free layer is then transferred to the interface while a fraction $(1-r)$ is transmitted to the reference layer spin accumulation. To model the rotation of the spin polarization of the spin current, we then assume that $-\Delta \mu^{\text{int}} / \tau_B$ is transferred from the interface to the free layer (see Fig.\ref{fig:mu_s_int}). This corresponds to a full rotation of the spin polarization. Considering a partial rotation would require, in our simple approach, to introduce an additional parameter which we wish to avoid for this qualitative modeling. In order to conserve angular momentum, however, we need to have a dissipation of the extra $2\Delta \mu^{\text{int}}/ \tau_B$ that is generated. Realistically, this dissipation should be independent of the thickness of the spacer layer (which is not the case here since $\tau_B$ is given by the spacer thickness divided by the Fermi velocity in the spacer). But this relationship is enforced by conservation of angular momentum and our assumption of full spin rotation upon reflection. Finally, all the spin current generated by the reference layer is assumed to be completely transmitted to the free layer without transiently stopping by the interface. The complete situation is summarized in Fig.\ref{fig:mu_s_int} and the corresponding spin current terms entering the spin accumulation dynamics equation (\ref{eq:dmu}) of each of the considered ferromagnetic layers are:
\begin{subequations}\label{eq:js}
\begin{gather}
\dfrac{\bm{\nabla}\cdot \bm{J}_s^{\text{free}}}{\overbar{D}} = \dfrac{\Delta \mu^{\text{free}}}{\tau_B} - \left( -\dfrac{\Delta \mu^{\text{int}}}{\tau_B} \right) - \dfrac{\Delta \mu^{\text{ref}}}{\tau_B}
\\[2ex]
\dfrac{\bm{\nabla}\cdot \bm{J}_s^{\text{ref}}}{\overbar{D}} = \dfrac{\Delta \mu^{\text{ref}}}{\tau_B} - (1-r)\left( \dfrac{\Delta \mu^{\text{free}}}{\tau_B} \right) \label{eq:js_ref}
\end{gather}
\end{subequations}
While the dynamics of the interfacial spin accumulation is:
\begin{equation}\label{eq:mu_s_int_dyn}
\dfrac{d \Delta \mu^{\text{int}}}{dt} = r\dfrac{\Delta \mu^{\text{free}}}{\tau_B} - \left( -\dfrac{\Delta \mu^{\text{int}}}{\tau_B} \right) - 2\dfrac{\Delta \mu^{\text{int}}}{\tau_B}
\end{equation}
Where we wrote all equations in such a way that each term appears in Fig.\ref{fig:mu_s_int}. By setting $r$ to zero, one retrieves ballistic spin transport as it is modeled by Beens \textit{et al.} \cite{Beens2020}. Overall, this model generalizes the model of reference \cite{Remy2023} and so it should also be suitable to reproduce ultrafast magnetization reversal provided that the calculated spin accumulation has the right dynamics and amplitude. For the simulations shown below, we use $r =$ 0.1 unless it is stated otherwise.

\begin{figure*}[t!]
\centering
\includegraphics[scale=1]{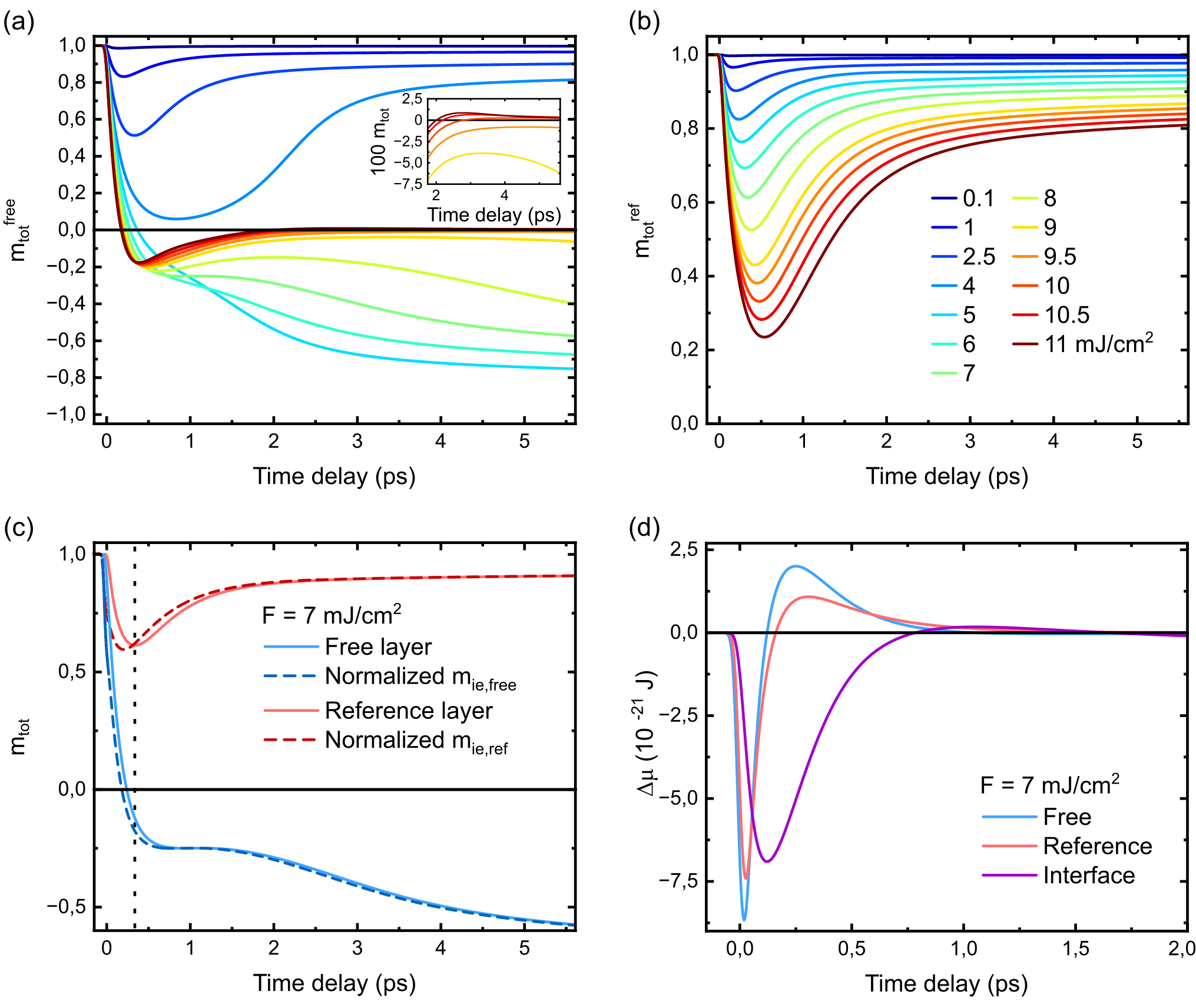}
\caption[P]{Normalized magnetization dynamics of the spin-valve sample where the dynamics of the free layer is shown in (a) and the one of the reference layer is shown in (b), for various fluences as indicated in (b). The inset of (a) shows a zoom between 2 and 5 ps where the normalized magnetization has been scaled by a factor of 100. (c) shows the same data as (a) and (b) for a fluence of 7 $\text{mJ/cm}^{\text{2}}$ together with the normalized instantaneous equilibrium $d$ electrons magnetization of each layer; the vertical dotted line indicates the time instant where the magnetization of the reference layer starts to recover. (d) shows the calculated spin accumulation dynamics in each layer as well as at the Cu/Reference layer interface.}\label{fig:P}
\end{figure*}

\subsection{P configuration}

\begin{figure*}[t!]
\centering
\includegraphics[scale=1]{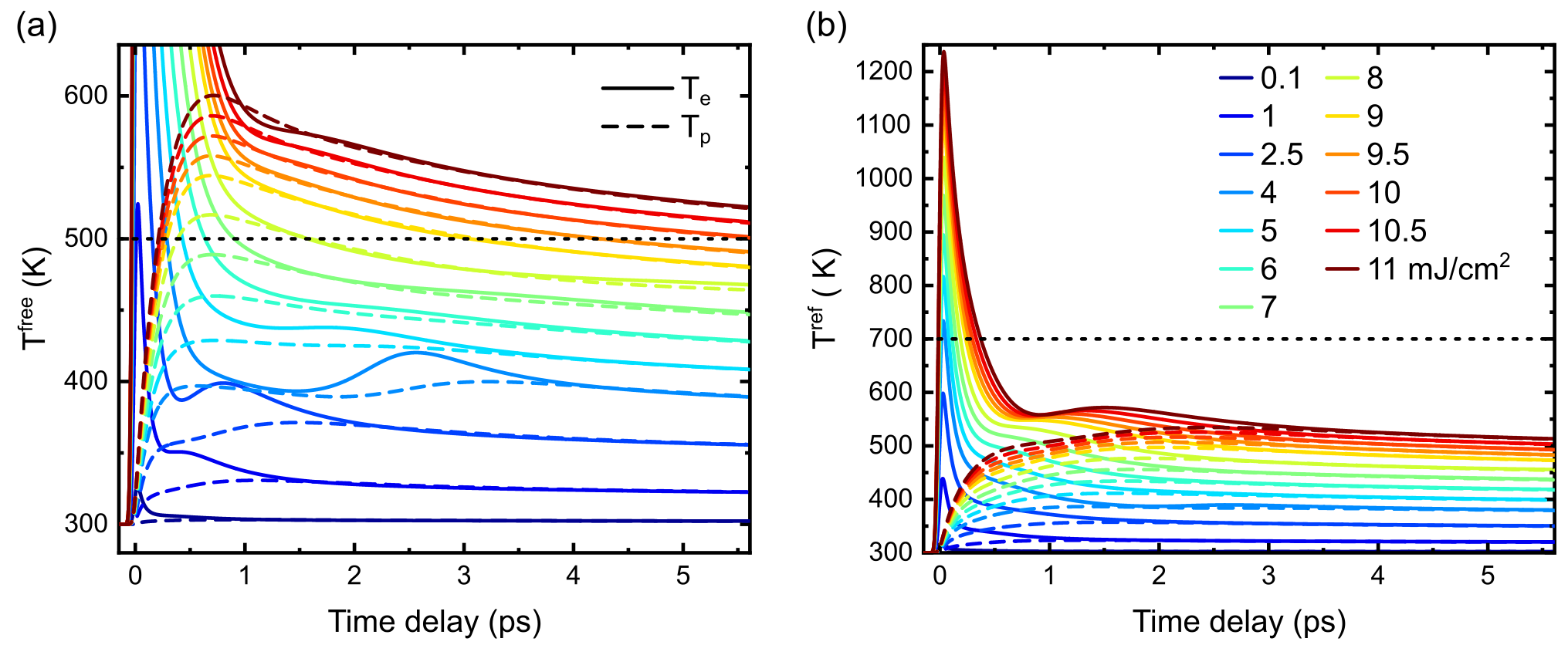}
\caption[T]{Temperature dynamics of electrons (solid lines) and phonons (dashed lines) for the (a) free and (b) reference layers. The dynamics is calculated for various fluences as indicated in (b). The Curie temperature of each layer is indicated in each case by the horizontal dotted line.}\label{fig:T}
\end{figure*}

In Fig.\ref{fig:P}, we show the results of the computed angular momentum dynamics in the spin-valve sample with our model. Fig.\ref{fig:P}(a) shows the normalized magnetization in the free layer while Fig.\ref{fig:P}(b) shows the normalized magnetization of the reference layer. The dynamics is computed for several fluences as shown in Fig.\ref{fig:P}(b). We see that the reference layer exhibits the standard UDM plus recovery behavior without any special feature. Consistently with Fig.\ref{fig:single_layer}, there is no CSD observable as the normalized magnetization never even reaches 0.2 for these fluences. We note however that the spin current coming from the demagnetization of the free layer hinders the demagnetization of the reference layer. This behavior is known to happen in real systems \cite{VanHees2020}. For the free layer, we observe only UDM plus recovery at low fluence, magnetization reversal above $F1 \sim$ 4 $\text{mJ/cm}^{\text{2}}$ and only a transient switching for fluences above $F2 \sim$ 10 $\text{mJ/cm}^{\text{2}}$. The transient nature of the latter switching is best observed in the inset of Fig.\ref{fig:P}(a). This behavior is consistent with the experimental observation of Igarashi \textit{et al.} \cite{Igarashi2023} that the free layer is only permanently reversed for a bounded range of fluences $[F1,F2]$ when starting from the P configuration. For fluences which are greater than $F2$ (yet still below the threshold fluence to generate a multidomain state due to a complete quenching of magnetization in the sample), the system remains in the P configuration on a long timescale. A key characteristic of the dynamics observed in reference \cite{Igarashi2023} is that the free layer magnetization crosses zero before the reference layer starts to remagnetize. The $s$-$d$ model of Beens \textit{et al.} cannot reproduce this feature \cite{Igarashi2023}. In Fig.\ref{fig:P}(c), we plot the magnetization dynamics of both layers for a fluence greater than $F1$ and lesser than $F2$. We also plot the instantaneous equilibrium magnetization of the $d$ electrons, for reference. We can see that with our model, the normalized magnetization of the free layer does cross zero before the reference layer starts to remagnetize (this instant is indicated by the vertical dotted line). The time delay between the free layer magnetization zero crossing and the beginning of the remagnetization of the reference layer increases with fluence. To provide more insight regarding the dynamics of angular momentum and spin transport, we plot the spin accumulations in each layer and at the Cu/Reference layer interface in Fig.\ref{fig:P}(d). We first note that the spin accumulation inside the free layer has the bipolar shape used in Ref. \cite{Remy2023}, together with the same order of magnitude, to obtain magnetization reversal of a [Co/Pt] multilayer subjected to a spin current coming from a ferrimagnetic GdFeCo. It is known that the spin accumulation needs to be positive to induce the reversal of magnetization. This happens naturally at lower fluences (see Fig.\ref{fig:single_layer}(d)) due to remagnetization. For fluences above $F1$, the electronic temperature overcomes the critical temperature of the free layer (see Fig.\ref{fig:T}(a)) and, without an external source of angular momentum, CSD will appear \cite{Remy2023}. From Eq.(\ref{eq:dmu}), it follows that the spin accumulation will be negative at all times (the terms containing the instantaneous equilibrium magnetization can lead to a positive spin accumulation when the light source term is not zero, but this effect is negligible). This ensures that a ferromagnetic layer cannot reverse its magnetization because of the spin accumulation it generates, in normal circumstances. The positive spin accumulation peak of the free layer in Fig.\ref{fig:P}(d) is due to the spin current reflection mechanism. One can see that after a certain delay due to the ballistic spin transport, a spin accumulation starts reaching the Cu/Reference layer interface. Because the corresponding angular momentum is reversed upon being reflected back to the free layer, this leads to an increase of the spin accumulation of the free layer which eventually becomes positive. Once the free layer spin accumulation becomes positive, the spin accumulation at the interface starts to decrease. The spin accumulation of the reference layer also has this bipolar structure because (i) the reference layer remagnetizes and (ii) the ballistic spin transport Eq.(\ref{eq:js_ref}) tends to bring the spin accumulation curve of the reference layer closer to the one of the free layer (and vice versa).

We now look at some other effects predicted by our model in this spin-valve. First, we look at the electron and phonon temperatures dynamics in each ferromagnetic layer. This is shown in Fig.\ref{fig:T}(a) and (b) for the free and reference layers respectively.
A standard dynamics, as obtained from the two temperature model, is obtained at first glance. Upon closer inspection, we can see however some bumps or additional peaks in the electron temperature dynamics for certain fluences. We adjusted the scale so as to make this effect obvious for the free layer, therefore cutting off the first peak of the dynamics which does not show any interesting feature. The deviation from the standard two temperature model dynamics is especially large when no reversal of magnetization is obtained. These extra features, such has the peak around 2,5 ps in the free layer for 4 $\text{mJ/cm}^{\text{2}}$ is due to the increase of magnetization (in absolute value) via the magnetization dependent term in Eq.(\ref{eq: Te}). Although these peaks may be overestimated due to the fact that the mean field model tends to overestimate the speed of remagnetization around $m = 0$, we still observe a sizable effect of the transfer of energy from $d$ to $s$ electrons in the reference layer while such effect did not exist for a single layer with parameters identical to that reference layer (Section \ref{sec:UDM}). We therefore conclude that an external source of angular momentum, via a spin current, can heat up electrons in a way that should be observable experimentally.

\begin{figure*}[t!]
\centering
\includegraphics[scale=1]{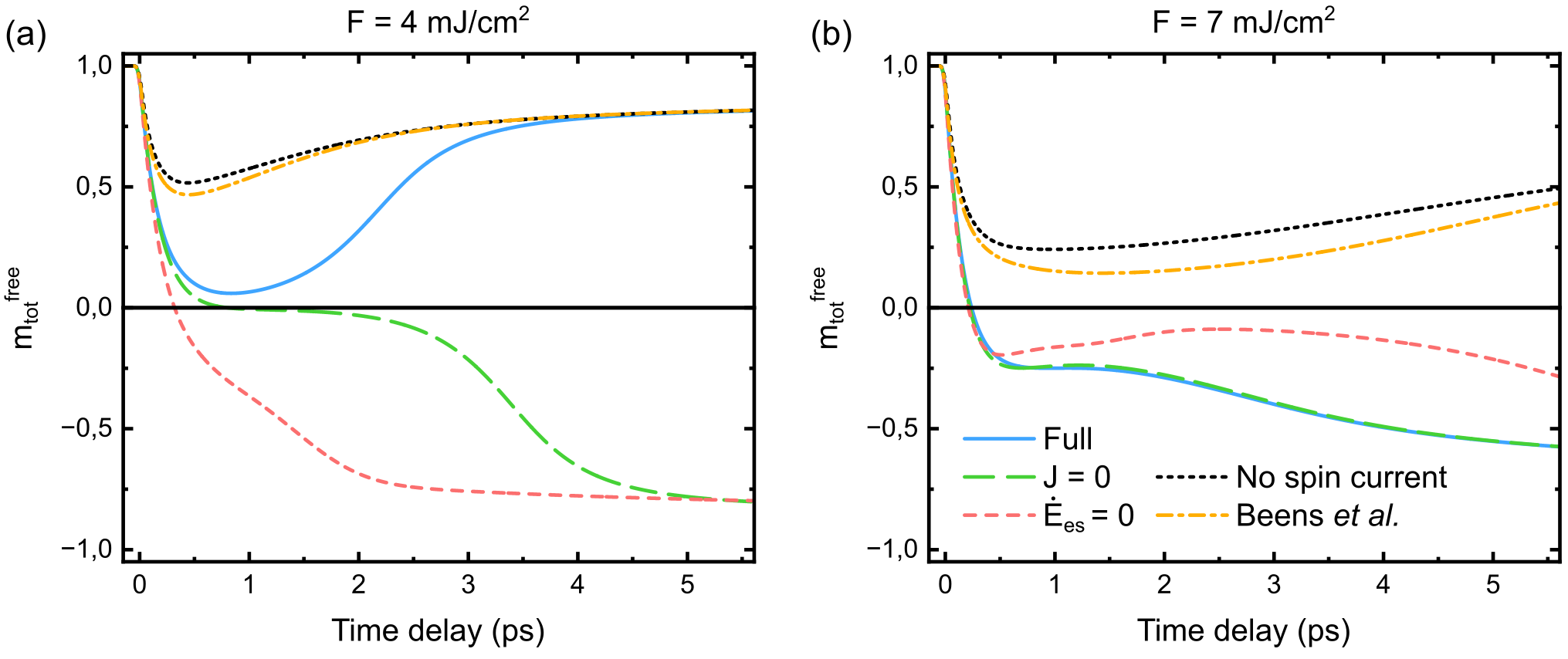}
\caption[Comp]{Normalized magnetization dynamics of the free layer computed for (a) 4 and (b) 7 $\text{mJ/cm}^{\text{2}}$ for different models. ``Full" corresponds to the full equation (\ref{eq:final_dmu}) with $J =$ 0.1 (reference layer) and 0.05 (free layer) eV and $\alpha =$ 1; ``$J =$ 0" is obtained by setting $J =$ 0 eV for both ferromagnetic layers; ``$\dot{E}_{\text{es}} =$ 0" is obtained by neglecting the magnetization dependent term in equation (\ref{eq: Te}); ``No spin current" is obtained by neglecting all spin transport; ``Beens \textit{et al.}" is obtained by setting $r =$ 0 and $J =$ 0.}\label{fig:Comp}
\end{figure*}

Now, we look at the role of the dynamic exchange splitting, spin currents and $d$ to $s$ energy transfer on the magnetization dynamics of spin-valves. To do so, we plotted in Fig.\ref{fig:Comp} the magnetization dynamics of the free layer for two fluences and different models. 
First, we can neglect the role of dynamic exchange splitting (``$J =$ 0"). For the highest fluence, where magnetization does not stay around zero, a very little difference is obtained. On the other hand, for the lower fluence, we can see a significant difference which results either in a reversal or not. Including the dynamic exchange splitting will not qualitatively change the magnetization dynamics but it can have sizable quantitative effects and so should also be considered in more realistic models. Then, we can also neglect the $d$ to $s$ electrons energy transfer (``$\dot{E}_{\text{es}} =$ 0"). This can lead to large effects but the reason is, just as for the single layer case, that it will change the maximum electron temperature. We can also block all spin transport which prevents any magnetization reversal even for high fluences (not shown). We can see that blocking this spin current also has a large effect on the magnetization dynamics. Even though our value of the reflection parameter $r$ is quite arbitrary, this should not come as a surprise as it was observed experimentally that spin heating (i.e. a decrease of magnetization solely due to an external source of angular momentum) can lead to a change of magnetization of up to 50$\%$ \cite{Remy2023}. Finally, we compare these results with the calculations obtained with the model of Beens \textit{et al.} \cite{Beens2020}. This model can also lead to magnetization reversal of the free layer but at higher fluence and the zero crossing happens for larger time delays \cite{Igarashi2023}. Because temperatures are higher for higher fluences, the maximum reachable normalized magnetization is smaller and the effect of the spin current in Beens \textit{et al.} model is smaller, at least for our choice of parameters. Overall, Fig.\ref{fig:P} and \ref{fig:Comp} illustrate the rich varieties of dynamics which can be expected in spin-valves compared to single layers where there is only demagnetization followed by remagnetization. This diversity and its strong fluence/parameters dependence should be kept in mind when studying the ultrafast magnetization dynamics of spin-valve heterostructures. Such diversity of behaviors has already been observed experimentally \cite{Igarashi2020, Remy2021}.

\subsection{AP configuration}

\begin{figure*}[t!]
\centering
\includegraphics[scale=1]{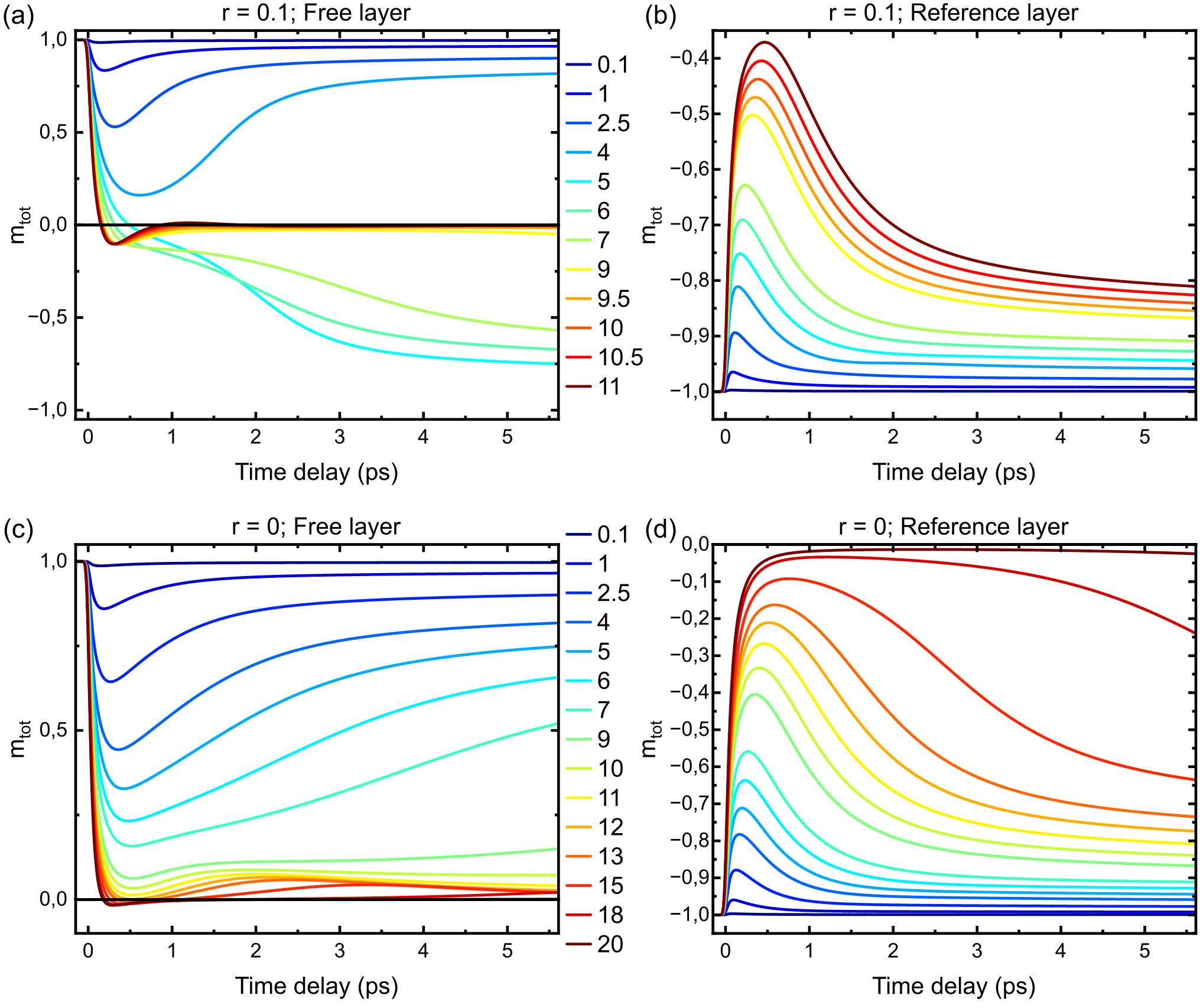}
\caption[AP]{Normalized magnetization dynamics for the (a), (c) free and (b), (d) reference layers. In (a) and (b), we use $r =$ 0.1 while in (c) and (d) we use $r =$ 0. For (a) and (b), the considered fluences (in $\text{mJ/cm}^{\text{2}}$) are indicated on the right of plot (a) while for (c) and (d) they are indicated on the right of plot (c).}\label{fig:AP}
\end{figure*}

To finish this section, we calculate the dynamics of the spin-valve when it is initially prepared in an AP configuration. In practice, we run exactly the same simulations except that out of the two stable equilibrium magnetizations of the reference layer, we select the solution with a negative sign instead of the positive one. We always keep the positive solution for the free layer. We assume that the reflection mechanism is identical to the P case. In particular, the spin polarization is rotated the same way independently of the magnetic configuration of the reference layer, consistently with the model introduced above. The results are shown in Fig.\ref{fig:AP}(a) and (b).
Apart from the obvious sign change for the reference layer, the results are almost identical to the simulations of the P case. A difference of interest, consistent with experiments \cite{Igarashi2023} is that it is harder to switch the free layer from the AP configuration compared to the P configuration, although the effect is much smaller than in the experiments (check the slightly reduced quenching of magnetization for a fluence of 4 $\text{mJ/cm}^{\text{2}}$ in both cases). However, in the experiments, the transient reversal of the free layer from the P configuration starts happening for fluences $F2$ almost identical to the threshold fluence $F1'$ required to observe a reversal of the free layer from the AP configuration i.e. $F1' \sim F2$. This is not the case in our simulations where $F1' \gtrsim$ 4 $\text{mJ/cm}^{\text{2}}$ while $F2 \sim$ 10 $\text{mJ/cm}^{\text{2}}$. We note that Igarashi \textit{et al.} \cite{Igarashi2023} do not consider this reflection mechanism for the AP configuration. We thus perform simulations for $r =$ 0 in Fig.\ref{fig:AP}(c) and (d). However, no permanent reversal of the free layer magnetization is observed. Although a transient reversal is observed starting from fluences around 12 $\text{mJ/cm}^{\text{2}}$, the remagnetization of the reference layer prevents the reversal from being permanent. Even at much higher fluences, when the remagnetization of the reference layer is greatly hindered due to CSD, a permanent reversal is still not possible. This is because the magnetization of the free layer is also further reduced which makes it even more sensitive to spin currents. In real systems, a multidomain state would be generated at such high fluences. The transient nature of the reversal at these higher fluences could also be due to the overestimation of remagnetization by mean field models. 

\begin{figure*}[t!]
\centering
\includegraphics[scale=1]{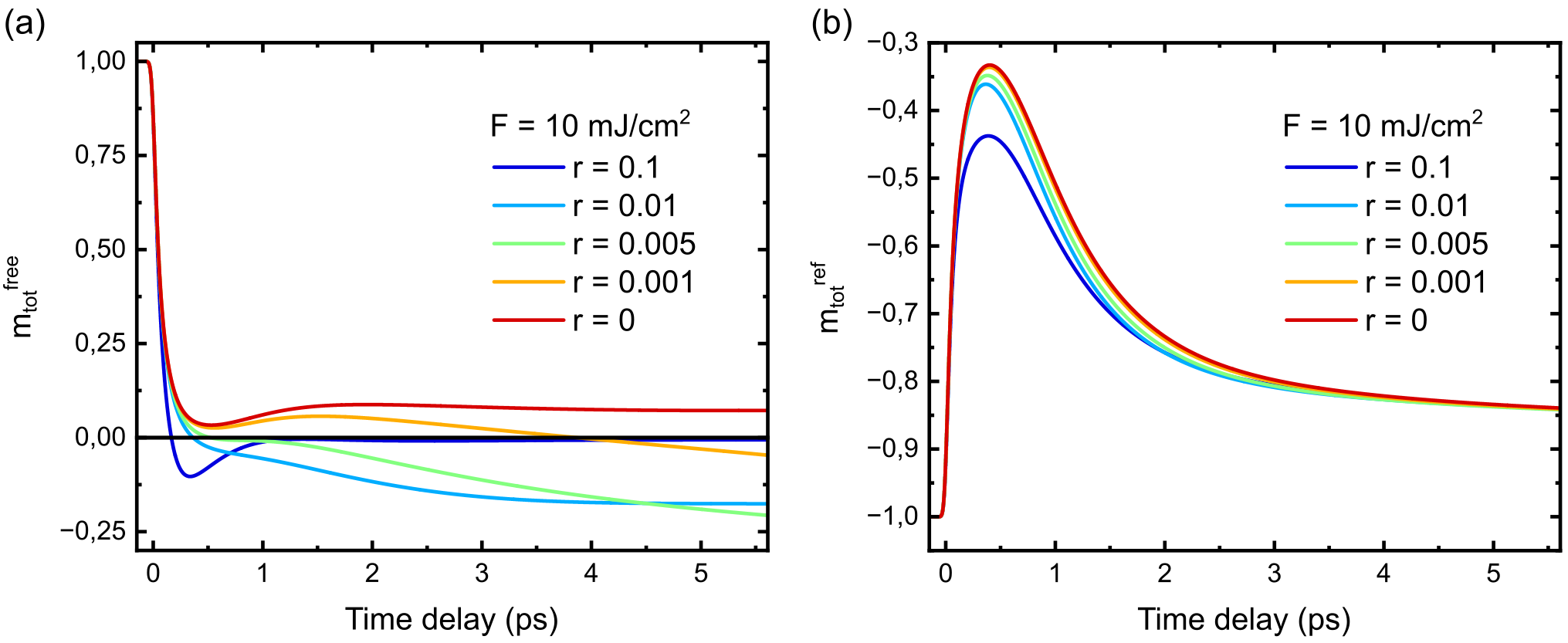}
\caption[APcomp]{Normalized magnetization dynamics, of the (a) free and (b) reference layers, in the AP configuration and for different values of $r$ as indicated.}\label{fig:APcomp}
\end{figure*}

It could also be possible that a reflection mechanism is still present in the AP configuration, even though it would be different, for reasons we do not explain, from the reflection mechanism when one starts from the P configuration. This is supported by some experimental measurements of Igarashi \textit{et al.} \cite{Igarashi2023} where a reversal of the free layer from an AP configuration is still observed for a copper spacer with a thickness of 40 nm. In this case, only a small amount of light can reach the reference layer which also supports a contribution of the spin current reflection mechanism. In Fig.\ref{fig:APcomp}, we compute the magnetization dynamics of both layers for a fluence around $F2$ and for different values of $r$.
The magnetization dynamics of the reference layer remains qualitatively the same for the considered reflection parameters. The one of the free layer is however significantly modified. As a general trend, we see that lowering the value of the $r$ parameter slows down the reversal dynamics i.e. the normalized magnetization crosses zero for larger time delays. It is also interesting to note that in all cases, the temperature dynamics are almost identical (plus or minus two Kelvins) even though the $d$ to $s$ electrons energy transfer is different for different values of $r$. This means that one cannot in general conclude that a magnetic subsystem is at equilibrium from the fact that its magnetization almost no longer changes (as it is observed for instance in Fig.\ref{fig:APcomp}(a) for $r =$ 0, 0.01 and 0.1). It is possible that an out of equilibrium situation is sustained due to a spin current emitted by another layer where the dynamics is still not over (here the reference layer; see Fig.\ref{fig:APcomp}(b)).

We conclude this section by highlighting that our spin current reflection mechanism is probably oversimplified. We note however that our approach, based on parameters with reasonable values, shows that this mechanism, if it exists, generates spin accumulations with a reasonable order of magnitude and triggers a magnetization dynamics with a realistic speed.

\section{\label{sec:conclusion}Conclusion}

In this work, we presented an extension of the $s$-$d$ model based on references \cite{Cywinski2007, Tveten2015, Gridnev2016, Beens2020} which includes a dynamic exchange splitting (and equilibrium spin polarization of the $s$ band), energy transfer from $d$ to $s$ electrons \cite{Gridnev2018, Griepe} as well as a newly proposed spin current reflection mechanism \cite{Igarashi2023}. This model leads to qualitatively (and also mostly quantitatively) identical results in single ferromagnetic layer systems compared to the case where all these additional effects are neglected. In the case where there is an external/non-local source of angular momentum such as in spin-valves, we showed that these effect can drastically change the observed magnetization dynamics. In particular, we could reproduce the magnetization reversal of the free layer of a ferromagnetic spin-valve which cannot be qualitatively reproduced by previous models. We also predict that the electronic temperature dynamics may be strongly affected, for instance in such spin-valves, when a ferromagnetic material is subjected to an external source of angular momentum.

The advantage of our model is that it is not computationally expensive but as pointed out in this paper, it lacks quantitative predictive power. In particular, the mechanism for the reflection of the spin current is still unclear. It is also not clear whether the $s$-$d$ model is well suited for such simulations as it does not contain magnons which are believed to play a fundamental role at the ferromagnetic/paramagnetic transition \cite{Scheid2023}. Transverse excitations however are included in this mean-field model, but the fast recovery of magnetization (we mean even in the absence of an external source of angular momentum), once it has almost been quenched, compared to experiments \cite{Remy2023} or atomistic simulations \cite{Kazantseva2008a}, seem to indicate that at least the mean-field approximation needs to be lifted in order to make more quantitative predictions. Still, our model is attractive for its simplicity and its capability to explain the diversity of ultrafast magnetization dynamics behaviors.

A quantitative agreement could also perhaps be obtained by following the approach of reference \cite{Griepe}, by using for instance temperature dependent parameters, but considering the number of parameters required in such simulations, it is not yet clear whether these calculations can have any quantitative predicting power for such complex systems. An alternative route, largely unexplored in the field of ultrafast magnetization dynamics, would be to treat the electron-magnon problem within the framework of the Fermi-liquid theory \cite{Korenman1977a}.

\appendix
\section{\label{sec: A3}Derivation of equation (\ref{eq: s_deltamu})}

In this appendix, we provide a derivation of Eq.(\ref{eq: s_deltamu}). This equation was first given in reference \cite{Tveten2015}, to the best of our knowledge, but its derivation and validity was not discussed. This derivation is useful to establish the limitations of the equations given in this work as well as to show the need for a more realistic description of the materials densities of state. We need to calculate the $s$ electrons spin polarization
\begin{equation}\label{eq: C_sz}
s^z = \dfrac{N_\uparrow-N_\downarrow}{2N}
\end{equation}
With $N = N_\uparrow+N_\downarrow$ the total number of $s$ electrons and $N_\sigma$ is the number of electrons with spin $\sigma$. Note that when all $s$ electrons have an up spin, $s^z = 1/2$ and so this definition of the spin polarization is consistent with the equations of the main text. The spin dependent electronic numbers are obtained from
\begin{equation}
N_\sigma = \int D_\sigma(E)f(E; T_e, \mu_\sigma)dE
\end{equation}
With $f$ the Fermi-Dirac distribution function. However, Eq.(\ref{eq: s_deltamu}) does not dependent on the electronic temperature explicitly. Thus, we first need to approximate the Fermi-Dirac functions by step functions. This means that the thermal energy should be much smaller than the width of the $s$ band: $k_BT_e \ll \mu_\sigma - E^0_\sigma$. Then
\begin{equation}\label{eq: N_DOS}
N_\sigma = \int_{E^0_\sigma}^{\mu_\sigma} D_\sigma(E-E^0_\sigma) dE = \mathcal{D}_\sigma(\mu_\sigma-E^0_\sigma) - \mathcal{D}_\sigma(0)
\end{equation} 
Where we shifted the functions representing the densities of state for convenience and we also assumed that these functions are continuous such that they all have an antiderivative $\mathcal{D}_\sigma$. We now perform three consecutive first order Taylor expansions of $\mathcal{D}_\sigma(\mu_\sigma-E^0_\sigma)$, assuming that $\Delta\mu$, $\mu - \mu_{\text{ie}}$ and $E^0_\sigma - E^0_{\sigma, \text{ie}}$ are small compared to $\mu_\sigma-E^0_\sigma$. $\mu_\sigma = \mu + \sigma ~ \Delta \mu /2$ and the ``ie" subscript refers to instantaneous equilibrium value as for the spin polarization $s^z_{\text{ie}}$. We obtain:
\begin{equation}
\begin{split}
N_\sigma &\simeq \mathcal{D}_\sigma(\mu_{\sigma,\text{ie}}-E^0_{\sigma,\text{ie}}) - \mathcal{D}_\sigma(0)\\
&+\sigma\dfrac{\Delta\mu}{2}D_\sigma(\mu-E^0_\sigma)\\
&+(\mu - \mu_{\text{ie}})D_\sigma(\mu_{\text{ie}}-E^0_\sigma)\\
&-(E^0_\sigma - E^0_{\sigma, \text{ie}})D_\sigma(\mu_{\text{ie}}-E^0_{\sigma,\text{ie}})
\end{split}
\end{equation}
To the same level of approximation, all the density of states factors should be taken equal, and after shifting back the function representing the densities of state, one has
\begin{equation}\label{eq: N_sig}
N_\sigma \simeq N_{\sigma,\text{ie}} + D_\sigma(\varepsilon_F)\left[ (\mu - \mu_{\text{ie}}) +\sigma\dfrac{\Delta\mu}{2} -(E^0_\sigma - E^0_{\sigma, \text{ie}}) \right]
\end{equation}
Using the fact that the total number of $s$ electrons does not change $N = N_{\text{ie}} = N_{\uparrow,\text{ie}}+N_{\downarrow,\text{ie}}$, one obtains:
\begin{equation}\label{eq: mu-muie}
\begin{split}
\mu - \mu_{\text{ie}} &= - \dfrac{D_\uparrow (\varepsilon_F) - D_\downarrow (\varepsilon_F)}{D_\uparrow (\varepsilon_F) + D_\downarrow (\varepsilon_F)} \dfrac{\Delta \mu}{2}\\ 
&+ \dfrac{D_\uparrow (\varepsilon_F)}{D_\uparrow (\varepsilon_F) + D_\downarrow (\varepsilon_F)} (E^0_\uparrow - E^0_{\uparrow, \text{ie}})\\ 
&+ \dfrac{D_\downarrow (\varepsilon_F)}{D_\uparrow (\varepsilon_F) + D_\downarrow (\varepsilon_F)} (E^0_\downarrow - E^0_{\downarrow, \text{ie}})
\end{split}
\end{equation}
Equation (\ref{eq: s_deltamu}) is readily obtained by combining equations (\ref{eq: C_sz}), (\ref{eq: N_sig}) and (\ref{eq: mu-muie}).

Now $(E^0_\sigma - E^0_{\sigma, \text{ie}}) \sim J$ and Eq.(\ref{eq: N_DOS}) implies that $\mu_\sigma-E^0_\sigma \sim 1/\overbar{D}$. So to be consistent, our theory should be such that $\Delta \mu$, $J$ and $k_BT_e$ are much smaller than $1/\overbar{D}$. We have used $J = 0.1/\overbar{D}$ and for the highest studied fluences, we have $k_BT_e \simeq 0.2/\overbar{D}$ and $\Delta \mu \simeq 0.1/\overbar{D}$. However, this does not invalidate the qualitative nature of our results since we obtain similar results for $J=0$ and the magnetization switching in the spinvalve appears at relatively low fluences. Nevertheless, we note that using larger values of $J$ (in combination with fluences comparable to the ones used in this work) leads to large divergences in the spin accumulation because of the factor $1/(1-JS\eta(t)/(2k_BT_e(t)))$ in Eq.(\ref{eq:final_dmu}). No such behavior was observed for the results presented in this work.

\begin{acknowledgments}
The author thank Jon Gorchon, Grégory Malinowski, Junta Igarashi, Guillermo Nava Antonio, Chiara Ciccarelli, Philippe Scheid and Stéphane Mangin for valuable discussions.

This work is supported by the ANR-20-CE09-0013 UFO, by the Institute Carnot ICEEL for the project “CAPMAT” and FASTNESS, by the Région Grand Est, by the Metropole Grand Nancy, for the Chaire PLUS by the impact project LUE-N4S, part of the French PIA project “Lorraine Université d’Excellence” reference ANR-15-IDEX-04-LUE, by the “FEDERFSE Lorraine et Massif Vosges 2014-2020”, a European Union Program, by the European Union’s Horizon 2020 research and innovation program COMRAD under the Marie Skłodowska-Curie grant agreement No 861300. This article is based upon work from COST Action CA17123 MAGNETOFON, supported by COST (European Cooperation in Science and Technology). We also acknowledge funding by the German Research Foundation (DFG) through the collaborative research center SFB TRR 227 “Ultrafast spin dynamics” (Project ID 328545488, project B02) and the European Union H2020 program through the FET project SKYTOP/Grant No. 824123.
\end{acknowledgments}

\bibliography{spin_acc_rev}

\end{document}